# TESTING THE BOUNDARY CONDITIONS OF GENERAL RELATIVITY NEAR THE EARTH-SUN SADDLE POINT[*]

Tom Martin
Gravity Research Institute
Boulder, Colorado 80306-1258
tmartin@rmi.net

## *Abstract*

*We suggest that a satellite with a stable atomic clock on board be sent through the Earth-Sun gravitational saddle point to experimentally determine whether Nature prefers static solutions of the field equations of General Relativity, such as the standard Schwarzschild solution, or whether Nature prefers equivalent non-static solutions. This is a test of the boundary conditions of General Relativity rather than of the field equations. The fractional difference in clock rates between the two possibilities is a part in $10^8$. This is a large and easily measurable effect.*

## Introduction

The standard Schwarzschild coordinates [1] are the static, spatially curved, and time dilating coordinates which are used to represent a simple spherically symmetric gravitating body in General Relativity. Most of the experimental verification of General Relativity rests on the Schwarzschild solution as represented by these coordinates. Given the popularity of this original form of the solution, it is rarely recognized that there is an alternative and equally valid *non-static* solution, obtained by a simple coordinate transformation, in which physical space may be interpreted as moving or flowing within the coordinate space instead of being static and curved within it. We will show that these *spatial flow coordinates* have a very simple physical interpretation. As long as we only examine the gravitational problem of a single body, the two representations are precisely equivalent as far as the physics goes (this is a result of the Principle of General Covariance). To express it another way, in the case of a single body, Nature seems not to discriminate between these static and non-static solutions. However, when we come to the *two body gravitational problem*, the spatially curving type coordinates and the spatially flowing type coordinates require distinctly different *boundary conditions* near the gravitational saddle point of the two bodies. It is in the region near the saddle point that Nature may best express a preference for one or the other types of these fundamental solutions.

---

[*] http://xxx.lanl.gov/ftp/gr-qc/papers/9806/9806033.pdf

Fortunately, the two different boundary conditions produce *highly distinct clock rates* near the saddle point. For the Earth-Sun saddle point, we will see that the fractional difference in clock rates for the two cases is almost a part in $10^8$. This means that the signal distinguishing the two cases will be four orders of magnitude above the noise in a simple cesium stabilized atomic clock whose nominal fractional frequency stability is a part in $10^{12}$. Better clocks will reveal even more of the structure around the saddle point. Any space mission can be used to distinguish between the two cases as long as the frequency of the clock on board can be externally monitored as it passes through the region of the saddle point.

## 1. Coordinates with Spatial Flow

For the purpose of discussion in this paper, we will be interested in global Galilean coordinates $\{\mathbf{r},t\}$ in which the proper time element of an atomic clock is given by

$$d\boldsymbol{t} = \boldsymbol{g}^{-1}\, dt, \tag{1-1}$$

where

$$\boldsymbol{g}^{-1} \equiv \sqrt{1 - u^2/c^2} \tag{1-2}$$

and

$$\mathbf{u} \equiv \mathbf{v} - \mathbf{w}. \tag{1-3}$$

Here, $\boldsymbol{t}$ is the proper time of the clock, $c$ is the coordinate speed of light in physical space (a constant), $t$ is the coordinate time, $\mathbf{r}$ is the position vector of the clock, $\mathbf{v} = d\mathbf{r}/dt$ is the coordinate velocity of the clock, $\mathbf{w} \equiv \mathbf{w}(\mathbf{r},t)$ is a 3-space vector field which we call *the flow of space*, and $\mathbf{u}$ is the velocity of the clock relative to physical space.

With this construct, we are obviously giving "space" an absolute physical significance. Newton's famous concept of "absolute space", in which no part of space is moving with respect to any other part, is characterized in this model by the condition that $\mathbf{w} \equiv \mathbf{0}$ everywhere. We are simply generalizing Newton's concept of absolute space to arbitrary spatial flows.

From the above equations, we see that the space-time line element is given by

$$c^2 d\mathbf{t}^2 = (c^2 - w^2)dt^2 + 2\mathbf{w} \cdot d\mathbf{r}\, dt - (d\mathbf{r})^2 \qquad (1\text{-}4)$$

in these Galilean coordinates with arbitrary flow $\mathbf{w}$.

In the following two Sections, we will show how the *canonical* General Relativistic examples of the rotating frame in flat space-time and the simple gravitational attractor of Schwarzschild in curved space-time can be cast into the form of Galilean frames with spatial flow.

## 2. The Rotating Frame in Flat Space-time

Let $\{x, y, z, t\}$ be a global rectangular Galilean coordinate system on flat space-time. This is an inertial Galilean frame with $\mathbf{w} = \mathbf{0}$, so its line element has the familiar form

$$c^2 d\mathbf{t}^2 = c^2 dt^2 - dx^2 - dy^2 - dz^2. \qquad (2\text{-}1)$$

If we let a second cylindrical Galilean frame $\{r, f, z, t\}$ rotate with rotational speed $\Omega$ counter-clockwise about the z-axis of the first frame, we will have induced in the cylindrical frame the cylindrically symmetric transverse spatial flow $\mathbf{w} = -\Omega r\, \hat{\mathbf{e}}_f$. Since $d\mathbf{r} = dr\, \hat{\mathbf{e}}_r + r df\, \hat{\mathbf{e}}_f + dz\, \hat{\mathbf{e}}_z$ and $\mathbf{w} \cdot d\mathbf{r} = -\Omega r^2 df$, expression (1-4) quickly gives us the line element in these non-inertial rotating coordinates:

$$c^2 d\mathbf{t}^2 = (c^2 - \Omega^2 r^2)dt^2 - 2\Omega r^2 df\, dt - (dr^2 + r^2 df^2 + dz^2). \qquad (2\text{-}2)$$

This is the usual General Relativistic solution for the rotating frame [2]. The proper time interval for a clock *at rest* in the rotating frame is seen from (2-2) to be given by

$$d\mathbf{t} = \sqrt{1 - \Omega^2 r^2 / c^2}\, dt\ . \qquad (2\text{-}3)$$

We will pause for a moment to consider the possible interpretations of this expression:

- From the point-of-view of the inertial frame $\{x, y, z, t\}$, the time dilation (2-3) is *Special Relativistic*, because the clock appears to be in circular motion with speed $v = \Omega r$. In this case, (2-3) is associated with the familiar Special Relativistic form $d\boldsymbol{t} = \sqrt{1 - v^2/c^2}\, dt$.

- From the point-of-view of the non-inertial frame $\{\boldsymbol{r}, \boldsymbol{f}, z, t\}$, the time dilation (2-3) is *General Relativistic*, and one might say something such as "the clock at rest is experiencing time dilation because it is immersed in the centrifugal potential $\Psi = -(1/2)\Omega^2 \boldsymbol{r}^2$". In this case, (2-3) is associated with the General Relativistic *Ansatz* $d\boldsymbol{t} = \sqrt{1 + 2\Psi/c^2}\, dt$ (this will be discussed below).

- From the point-of-view of *spatial flow*, however, the General Relativistic time dilation (2-3) of the clock at rest in the rotating frame is due to the clock's *immersion* in the spatial flow $\mathbf{w} = -\Omega r\, \hat{\mathbf{e}}_f$. Then, (2-3) is associated with the spatial flow form $d\boldsymbol{t} = \sqrt{1 - w^2/c^2}\, dt$ (which is derived from (1-4) for a clock at rest).

The imaginative reader may begin to see from all of this how the spatial flow point-of-view might reveal the essence of the Principle of Equivalence in General Relativity, but we do not pursue this line of reasoning here. For now, it is important to recognize that, from the point-of view of the rotating Galilean frame, it is the non-zero transverse circulating *flow* $\mathbf{w} = -\Omega r\, \hat{\mathbf{e}}_f$ which is responsible for slowing the clock.

## 3. The Schwarzschild Solution

We now show how it is possible to cast the Schwarzschild solution into the form of a Galilean frame with spatial flow. This will also provide us with a second example of the slowing of clocks by spatial flow.

Consider the flow

$$\mathbf{w} = \sqrt{2GM/r}\ \hat{\mathbf{e}}_r \qquad (3\text{-}1)$$

in a spherical Galilean coordinate frame $\{r, \boldsymbol{q}, \boldsymbol{f}, t\}$ centered on a simple spherically symmetric gravitational attractor of mass $M$. ($G$ is the gravitational constant.) This flow is radially directed *outward* from the attractor with a speed that is equivalent to the speed of free-fall from rest at infinity in ordinary Newtonian gravitational theory. (One could also use an inwardly directed flow of the same magnitude, as can be seen in the development below.)

Since $d\mathbf{r} = dr\,\hat{\mathbf{e}}_r + r\,d\theta\,\hat{\mathbf{e}}_\theta + r\sin\theta\,d\phi\,\hat{\mathbf{e}}_\phi$, we see that the Galilean line element (1-4) takes the form

$$c^2 d\tau^2 = c^2(1 - 2GM/rc^2)\,dt^2 + 2\sqrt{2GM/r}\;dr\,dt - (dr^2 + r^2 d\theta^2 + r^2 \sin^2\theta\,d\phi^2)\,. \quad (3\text{-}2)$$

We observe that this line element shows us that a clock at rest in the Galilean frame and immersed in the outwardly directed flow (3-1) experiences the usual General Relativistic gravitational time dilation

$$d\tau = \sqrt{1 - 2GM/rc^2}\;dt\;. \quad (3\text{-}3)$$

From our point-of-view, it is the non-zero radial *flow* (3-1) which slows the clock.

In fact, the Galilean frame with spatial flow (3-1) gives us *all of the correct General Relativistic physical effects* usually associated with the standard static and curved space Schwarzschild solution. We know this, because in General Relativity we are free to use any coordinate system we choose (Principle of General Covariance) [3], and if we de-synchronize Galilean coordinate time with the infinitesimal coordinate transformation (leaving the spatial coordinates unchanged)

$$dt_S \equiv dt + (1 - 2GM/rc^2)^{-1}\sqrt{2GM/r}\;c^{-2} dr\,, \quad (3\text{-}4)$$

the line element (3-2) transforms into the *standard Schwarzschild line element* [1],

$$c^2 d\tau^2 = c^2(1 - 2GM/rc^2)\,dt_S^2 - (1 - 2GM/rc^2)^{-1} dr^2 - (r^2 d\theta^2 + r^2 \sin^2\theta\,d\phi^2)\,. \quad (3\text{-}5)$$

In other words, the standard Schwarzschild coordinates $\{r,\theta,\phi,t_S\}$ for a spherically symmetric attractor in General Relativity are simply *de-synchronized Galilean coordinates*. Since the physics of orbits (Mercury's perihelion), clocks (gravitational time dilation), light paths (bending of light), and so on is always coordinate invariant in General Relativity (Principle of General Covariance), and since the Galilean coordinates are related to the Schwarzschild coordinates by a simple transformation, we are guaranteed that the spatial flow (3-1) gives us a solution which satisfies Einstein's field equations and gives all of the usual General Relativistic effects associated with the attractor of mass $M$. So, instead of claiming that the solution of Einstein's equations for a simple attractor is a solution with static space curvature (Schwarzschild solution), we can claim with equal validity that the solution is one that *exudes space* (Galilean solution). Einstein's equations predict that matter expels space. (As mentioned above, it is also possible that matter absorbs space.) This flow is no less a theoretical possibility than is the curvature of space.

The boundary conditions relevant to the Schwarzschild solution are that the coordinates should become Minkowskian (i.e., flat) at large distances from the attractor. In our Galilean spatial flow coordinates, this translates into the vanishing of **w** at large distances.

Now, in consideration of what we have shown, when we ask ourselves if there is any way we can experimentally determine whether Nature prefers one or the other forms of these solutions (static curvature or spatial flow), the first idea that occurs to us is that the region around the saddle point of the two body gravitational problem might be some sort of region of quiescence for the flow. We will address this question in Section 4.

The Stokes-Helmholtz vector decomposition theorem tells us that an arbitrary vector field can be written as a unique sum (up to a constant vector) of a solenoidal (divergence free) vector field and an irrotational (curl free) vector field. A constant spatial flow **w** is called an inertial flow. A purely solenoidal flow (the rotational flow of Section 2 is the canonical example) is called a purely non-inertial flow. A purely irrotational flow (the example of this Section is the canonical example) is called a purely gravitational flow.

Once we have introduced the generalization of Newton's absolute space as arbitrary spatial flow into physics, with the gravitational attractors expelling (or possibly absorbing) space, we can re-examine the foundations of physics from an entirely new perspective. These issues will be discussed in subsequent publications.

In concluding this Section, we mention that the total time dilation of a clock in arbitrary motion in an arbitrary spatial flow can be very easily calculated. Equation (1-1) takes the form

$$d\boldsymbol{t} = \sqrt{1 - (v^2 + w^2 - 2\mathbf{v}\cdot\mathbf{w})/c^2}\, dt \; . \tag{3-6}$$

For example, a satellite in circular motion with angular speed $\Omega$ centered on a spherically symmetric non-spinning planet experiences the total time dilation

$$d\boldsymbol{t} = \sqrt{1 - \Omega^2 \boldsymbol{r}^2/c^2 - 2GM/\boldsymbol{r}c^2}\ dt, \tag{3-7}$$

where $\boldsymbol{r}$ is the Galilean coordinate distance from the planet's center to the satellite. From this Galilean point-of-view, the time dilation is due to the interaction of the satellite with physical space.

A person solving this satellite problem from the point-of-view of the Schwarzschild coordinates would obtain essentially the same result, because (3-4) shows us that Schwarzschild coordinate time and Galilean coordinate time are approximately syntonous ($dt_S \cong dt$) when $\sqrt{2GM/r} \ll c$. He might interpret the result as allowing him to add the centrifugal and gravitational Newtonian potentials according to the "sum of the potentials" *Ansatz*

$$d\boldsymbol{t} \cong \sqrt{1 + 2(\Psi_{CENTRIFUGAL} + \Psi_{GRAVITATIONAL})/c^2}\ dt_S. \tag{3-8}$$

## 4. The Gravitational Two Body Problem

We now address the question of how to use the region around the Earth-Sun gravitational saddle point to experimentally determine whether Nature has a preference for either the static curved space type solutions or else for the non-static and spatially flowing type solutions of Einstein's field equations.

As is well known, the two body problem in General Relativity has no known analytical solution (this is analogous to the case in Newtonian gravitational physics in which the finite three body problem has no analytical solution). Never-the-less, there is a conventional approximate weak field solution of the two body problem in General Relativity [4] which is based on the *conventional boundary conditions* that the metric and its derviatives are continuous and that the metric is asymptotically flat (i.e., Minkowskian) at great distances from the two attractors. To first order, the solution for two bodies [5] is the same as the Newtonian two body solution obtained by adding the Newtonian potentials.

Consider the simple case of two attractors of equal mass: $M_1 = M_2$. At the midpoint between the two masses, the weak field solution with its conventional boundary conditions predicts [4,5] that, to first order, a clock at rest will experience the time dilation

$$d\boldsymbol{t} \cong \sqrt{1 - 2GM_1/\boldsymbol{r}c^2 - 2GM_2/\boldsymbol{r}c^2}\ dt_S\ , \qquad (4\text{-}1)$$

where $\boldsymbol{r}$ is the coordinate distance from the center of each mass to the midpoint, and $t_S$ is the Schwarzschild-like time coordinate for the weak field solution. This fits in with the "sum of the potentials" *Ansatz* for static and spatially curved type solutions mentioned in Section 3:

$$d\boldsymbol{t} \cong \sqrt{1 + 2(\Psi_1 + \Psi_2)/c^2}\ dt_S\ . \qquad (4\text{-}2)$$

On the other hand, from the point-of-view of a non-static and spatially flowing solution of the two body problem, there must be a well-defined spatial flow $\mathbf{w}$ everywhere. The symmetry of the case of two equal masses (both of which are assumed to be expelling space equally) requires the flow to be quiescent at the midpoint between the masses. This involves a *new set of boundary conditions* which at least require that the flow must vanish at the midpoint and also at great distances from the attractors. The vanishing of the spatial flow at the midpoint of the masses tells us, by means of equation (3-6), that a clock at rest there will experience *no* gravitational time dilation. Contrast this with the predicted time dilation of the static and curved space type solution (4-1).

Consider next the case of unequal masses such as the Earth-Sun system (we neglect lunar and other planetary perturbations for the sake of discussion). If $M$ and $R$ denote the mass of the Sun and its distance from the saddle point, and if $m$ and $r$ denote the mass of the Earth and its distance from the saddle point, then we will have approximately

$$r/R \cong \sqrt{m/M}\ . \qquad (4\text{-}3)$$

This locates the saddle point of the Earth-Sun system at about 260,000 kilometers from the Earth. Of course, the Earth-Sun system is rotating about its barycenter, and hence the saddle point is also rotating. The weak field solution with its conventional boundary conditions predicts [4,5] that, to first order, and from the point-of-view of a non-rotating Schwarzschild-like coordinate system centered on the barycenter, a clock comoving with the saddle point will experience the time dilation

$$dt_{CONVENTIONAL} \cong \sqrt{1 - \Omega^2 r^2/c^2 - 2GM/Rc^2 - 2Gm/rc^2} \; dt_S \;, \qquad (4\text{-}4)$$

where $r$ is the distance from the barycenter to the saddle point and $\Omega$ is the angular speed of the saddle point (we are assuming approximately circular orbits). Again, the "sum of the potentials" *Ansatz* is fulfilled.

Since we have no exact conventional analytical solution of the two body problem, we have no way to extract an exact analytical spatial flow type solution as we did in the case of the Schwarzschild solution for a single body. But we do have significant information from the requisite difference in boundary conditions. From the point-of-view of a non-rotating Galilean coordinate system centered on the barycenter, the spatial flow **w** must vanish at the saddle point. In this frame of reference, a clock comoving with the saddle point has speed $v = \Omega r$. Equation (3-6) gives us the time dilation for this comoving clock:

$$dt_{FLOW} = \sqrt{1 - \Omega^2 r^2/c^2} \; dt \;. \qquad (4\text{-}5)$$

The fractional difference in clock rates for the two possible sets of boundary conditions and their corresponding time dilations (4-4) and (4-5) is given approximately by

$$\sqrt{1 - \Omega^2 r^2/c^2} - \sqrt{1 - \Omega^2 r^2/c^2 - 2GM/Rc^2 - 2Gm/rc^2}$$
$$\cong GM/Rc^2 + Gm/rc^2 \cong 10^{-8} \qquad (4\text{-}6)$$

for the Earth-Sun system. This is a large and easily measurable difference.

## 5. Conclusion

We have shown that there remains a major experimental test of General Relativity. It is a test of the *boundary conditions* rather than of the field equations. Nothing could be simpler as a space mission testing fundamental physics than a *probe* sent directly through the Earth-Sun saddle point. All that is required is that *any* mission be sent through this point with a sufficiently stable clock on board whose frequency can be monitored throughout the passage. As mentioned in our introduction, a typical cesium stabilized atomic clock would be sufficient. Of course, it would be advantageous to have a *dedicated* mission with a highly stabilized clock on board which could be used to explore the time dilation structure of the entire region around the saddle point.